\input epsf.tex
\documentclass[12pt]{article}
\usepackage{graphicx}
\csname @addtoreset\endcsname{equation}{section}

\def\bseq{\begin{subequation}}  
\def\eseq{\end{subequation}}
\def\bsea{\begin{subeqnarray}}  
\def\esea{\end{subeqnarray}}

\newcommand{\bbox}{\lower.2ex\hbox{$\Box$}}

\newcommand{\beq}{\begin{equation}}
\newcommand{\eeq}{\end{equation}}
\newcommand{\bea}{\begin{eqnarray}}
\newcommand{\eea}{\end{eqnarray}}
\newcommand{\ena}{\end{eqnarray}}

\renewcommand{\a}{\alpha}
\renewcommand{\b}{\beta}

\renewcommand{\d}{\delta}

\newcommand{\pa}{\partial}
\newcommand{\g}{\gamma}
\newcommand{\G}{\Gamma}

\newcommand{\e}{\epsilon}

\renewcommand{\l}{\lambda}
\renewcommand{\L}{\Lambda}
\newcommand{\m}{\mu}

\newcommand{\n}{\nu}

\newcommand{\p}{\pi}

\newcommand{\s}{\sigma}

\newcommand{\Db}{\bar{D}}

\newcommand{\ad}{{\dot{\alpha}}}
\newcommand{\bd}{{\dot{\beta}}}

\begin{document}
\begin{titlepage}
\begin{flushright}
IFUM-FT-665\\

\end{flushright}
\vspace{1cm}
\begin{center}
{\Large \bf One-loop four-point function in\\
\vspace{3mm}

 noncommutative ${\cal N}
=4$ Yang-Mills theory}
\vfill
{\large \bf
Alberto Santambrogio ~and~
Daniela Zanon}\\
\vskip 7mm
{\small
Dipartimento di Fisica dell'Universit\`a di Milano
and\\ INFN, Sezione di Milano, Via Celoria 16,
20133 Milano, Italy\\}
\end{center}
\vfill
\begin{center}
{\bf Abstract}
\end{center}
{\small We compute the one-loop four-point function
in ${\cal N}=4$
supersymmetric Yang-Mills theory with gauge group $U(N)$.
We perform the calculation in ${\cal N}=1$ superspace using the
background field method and obtain the complete off-shell contributions to
the effective
action from planar and non planar supergraphs. In the low-energy
approximation the result simplifies and we can
study its properties under gauge transformations. It appears
that the nonplanar contributions
do not maintain the gauge invariance of the classical action.}
\vspace{2mm} \vfill \hrule width 3.cm
\begin{flushleft}
e-mail: alberto.santambrogio@mi.infn.it\\
e-mail: daniela.zanon@mi.infn.it
\end{flushleft}
\end{titlepage}

\section{Introduction}

The realization that the field theory limit of open strings in the
presence of a constant $B$ field gives rise to noncommutative gauge
theories has driven much attention on the subject \cite{all,SW,others}.
Several computations have been performed in the string amplitude
framework \cite{string} as well as in the perturbative field theory
one \cite{perturb}.
In dealing with supersymmetric theories it is often advantageous to
use a superspace formulation. Such an approach has been introduced
also for noncommutative theories \cite{FL,susy,schweda,DZ}. The net result is
that one constructs the theory in terms of standard superfields and
implements the non commutativity through the $*$-product which
replaces the ordinary multiplication. Since the $*$-product does not
affect the fermionic coordinates, the quantization and the
derivation of the Feynman rules can be performed following the same
procedure as for commutative theories.
Exponential factors in the interactions, stemming from the
$*$-products between the superfields, are essentially the only new feature
which distinguishes the noncommutative perturbation theory from
the commutative one. Indeed it has been shown that one
can use supergraph techniques and standard $D$-algebra both for
chiral matter models \cite{schweda} and for super gauge theories
\cite{DZ}.

In this paper we continue the work presented in \cite{DZ}.
There the noncommutative
${\cal N}=4$ supersymmetric Yang-Mills theory with gauge group $U(N)$
has been studied
in ${\cal N}=1$ superspace and  the background field quantization has
been derived  for  one-loop calculations in perturbation
theory. Here we follow the ideas and techniques introduced in \cite{DZ} and
compute the complete four-point function with external gauge fields
at one loop. We analyze the various contributions to the
effective action from planar and
nonplanar supergraphs. We find that our result is in accordance with the
off-shell extension of the result
from one-loop four gluon scattering on
parallel $D3$-branes \cite{*trek}. As noticed in \cite{*trek}, while
the on-shell $S$-matrix calculation is gauge invariant, the same is
not true for the off-shell contribution to the effective action. The
terms that do not respect the noncommutative gauge invariance of the
classical action are the ones produced by nonplanar diagram
contributions. At the end of the paper we discuss this issue.

The next section contains a brief
introduction to the background field method and the derivation of the
Feynman rules relevant for one-loop calculations. We also give
details on how we organize the Wick expansion and pay  special attention
to the various steps necessary for the construction of the nonplanar
diagrams. In section $3$ and $4$ we give the results for the planar
and nonplanar contributions. In section $5$ we study the
four-point function in
the low-energy approximation. Finally we conclude with comments on
the apparent loss of gauge invariance of the one-loop result and on
possible attempts to resolve the problem.

\section{Background covariant quantization for noncommutative ${\cal
N}=4$ Yang Mills}

The $*$-product is the  object of primary interest in
noncommutative field theories. Such an operation has been introduced
 also in
the context of noncommutative supersymmetric theories. It acts as a
non local multiplication for chiral and gauge superfields
in the following way
\beq
(\phi_1 * \phi_2)(x,\theta,\bar{\theta})\equiv e^{\frac{i}{2}
\Theta^{\m\n}\frac{\pa}{\pa x^\m}\frac{\pa}{\pa y^\n}}~
\phi_1(x,\theta,\bar{\theta})\phi_2(y,\theta,\bar{\theta})|_{y=x}
\label{starprod}
\eeq
It  maintains  explicit
supersymmetry and it can be used consistently in the construction of
noncommutative supersymmetric actions.

In terms of ${\cal N}=1$ superfields the classical action
of the noncommutative
${\cal N}=4$ supersymmetric Yang-Mills theory
can be written as (we use the
notations and conventions adopted in \cite{superspace,DZ})
\bea
&&S= \frac{1}{g^2}~{\rm Tr} \left( \int~ d^4x~d^4\theta~ e^{-V}
\bar{\Phi}_i e^{V} \Phi^i +\frac{1}{2} \int ~d^4x~d^2\theta~ W^2
+\frac{1}{2} \int ~d^4x~d^2\bar{\theta}~ \bar{W}^2 \right.\nonumber\\
&&\left.\left.~~~~~~~~~~~~~+\frac{1}{3!} \int ~d^4x~d^2\theta~ i\e_{ijk}
 \Phi^i
[\Phi^j,\Phi^k] + \frac{1}{3!}\int ~d^4x~d^2\bar{\theta}~ i\e^{ijk} \bar{\Phi}_i
[\bar{\Phi}_j,\bar{\Phi}_k] \right)\right|_*
\label{SYMaction}
\eea
where the symbol $|_*$ denotes multiplication  as defined in (\ref{starprod}).
In (\ref{SYMaction}) the $\Phi^i$ with $i=1,2,3$ are three chiral
superfields, while $W^\a= i\bar{D}^2(e_*^{-V}*D^\a e_*^V)$ is the gauge
superfield strength. All the fields are Lie-algebra valued, e.g.
$\Phi^i=\Phi^i_a T_a$, in the adjoint representation of $U(N)$.

The background field quantization  has been used efficiently in perturbative
calculations for commutative SYM theory
\cite{GS,superspace,GZ}.
In \cite{DZ} it has been shown that the superspace background field
method can be applied in the quantization procedure even for
noncommutative  gauge theories. This approach simplifies the
calculations dramatically. At one loop the four-point vector
amplitude receives contributions only from the vector superfields
themselves, since the ghost loops exactly
cancel the chiral matter loops. Moreover, as
in ordinary commutative gauge theories, the background field method
allows to express the one-loop corrections to the action in (\ref{SYMaction})
in terms of field strengths.

The background field quantization can be implemented in the
noncommutative theory essentially because the $*$-product does not
affect the superspace fermionic coordinates and therefore does not
alter the various properties of the superfields. In all the steps it
is sufficient to introduce the appropriate $*$-multiplication \cite{DZ}.
Here we briefly summarize how one performs the quantum background
splitting.
One defines covariant derivatives
\beq
\nabla_\a =e_*^{-\frac{V}{2}}* D_\a~ e_*^{\frac{V}{2}}\qquad
\qquad \bar{\nabla}_\ad=e_*^{\frac{V}{2}}* \bar{D}_\ad ~e_*^{-\frac{V}{2}}
\label{covder}
\eeq
so that the gauge Lagrangian becomes
\beq
\frac{1}{2}{\rm Tr}~ W^\a*W_\a=
-{\rm Tr} \left( \frac{1}{2}[\bar{\nabla}^\ad,\{\bar{\nabla}_\ad,
\nabla_\a\}]\right)^2
\label{lagrang}
\eeq
The splitting is obtained via the introduction of
a quantum prepotential $V$ and background covariant
derivatives
\beq
\nabla_\a ~\rightarrow~e_*^{-\frac{V}{2}}*\nabla_\a ~ e_*^{\frac{V}{2}}\qquad
\qquad \bar{\nabla}_\ad~\rightarrow~
e_*^{\frac{V}{2}}* \bar{\nabla}_\ad ~e_*^{-\frac{V}{2}}
\label{backcovder}
\eeq
On the r.h.s. of (\ref{backcovder})
the covariant derivatives are expressed in terms of
background connections, i.e.
\beq
\nabla_\a= D_\a-i{\bf{\G}}_\a \qquad \qquad \bar{\nabla}_\ad= \bar{D}_\ad-
i\bar{\bf{\G}}_\ad \qquad\qquad \nabla_a=\pa_a-i{\bf{\G}}_a
\label{backcovderconn}
\eeq
Adding to the classical Lagrangian (\ref{lagrang}) background covariantly
chiral gauge fixing functions, $\nabla^2 V$ and $\bar{\nabla}^2 V$.
one obtains
\beq
-\frac{1}{2g^2} {\rm Tr}\left[ \left(e_*^{-V}*\nabla^\a e_*^V\right)*
\bar{\nabla}^2  \left(e_*^{-V}*\nabla_\a e_*^V\right)
+ V*\left(\nabla^2\bar{\nabla}^2+\bar{\nabla}^2
\nabla^2\right)V\right]
\eeq
We are interested in one-loop calculations, therefore we can restrict
our attention to the terms in the action which are quadratic in the
quantum fields.
With the connections defined
in (\ref{backcovderconn}) we obtain
\bea
&&-\frac{1}{2g^2} {\rm Tr} ~V*\left[\frac{1}{2} \pa^a\pa_a-i{\bf \G}^a *\pa_a
-\frac{i}{2} \pa^a{\bf\G}_a * -\frac{1}{2}{\bf\G}^a *{\bf
\G}_a * \right.\nonumber\\
&&\left.~~~~~~~~~~~~~~~ -i {\bf W}^\a * (D_\a-i{\bf\G}_\a *)-i
 \bar{{\bf W}}^\ad
*( \bar{D}_\ad-i\bar{{\bf \G}}_\ad * )\right] V
\label{oneloopaction}
\eea
We have interactions with
the background fields at most linear in the $D$'s.
 Since
superspace Feynman rules require the presence of two $D$'s and
two $\bar{D}$'s for a non
zero loop contribution, at least
four vertices are needed and the first non vanishing correction
to the effective action is at the
level of the four-point function.
Thus following \cite{DZ}, we consider
\beq
-\frac{1}{2g^2} {\rm Tr} ~V*\left[\frac{1}{2} \pa^a\pa_a
-i {\bf W}^\a * D_\a -i \bar{{\bf W}}^\ad *
 \bar{D}_\ad\right] V
\label{relevantoneloopaction}
\eeq
and read from here
the superspace Feynman rules which, with the appropriate
$*$-product inserted, are the standard ones.
In momentum space
the vector propagators are
\beq
<V^a(\theta)V^b(\theta')>=-\frac{g^2}{p^2}\d^{ab} \d^4(\theta-\theta')
\label{prop}
\eeq
and the relevant interactions with the background are given in
(\ref{relevantoneloopaction}).
As already discussed in \cite{DZ}, one-loop contributions with
chiral matter fields and ghosts inside the loop cancel among
themselves, so we will consider only quantum vector loops.

In momentum space, with momenta $k_1, k_2, k_3$ flowing into the vertex
($k_1+k_2+k_3=0$), the three-point interactions can be written as
\bea
&&\frac{1}{2g^2}~\left({\cal U}(k_1,k_2,k_3) +\bar{\cal U}(k_1,k_2,k_3)
\right)\equiv\nonumber\\
&&~~~~~~~~~~~\equiv \frac{1}{2g^2}V_a(k_1)\left[~i {\bf W}_b^\a(k_2) D_\a+i
\bar{{\bf W}}_b^\ad(k_2)
\bar{D}_\ad~\right] V_c(k_3)~\nonumber\\
&&~~~~~~~~~~~~~~~~~~~~~~~~~~~~~~~~{\rm Tr}(T^aT^bT^c)~ e^{-\frac{i}{2}(k_1
\times k_2+k_2\times k_3+k_1\times k_3)}
\label{vertex}
\eea
where $k_i\times k_j\equiv
(k_i)_\m\Theta^{\m\n}(k_j)_\n$.
Inserting a vertex  in the loop we obtain two types of terms, an
{\em untwisted} and a {\em twisted} term, i.e.
\bea
&&{\cal U}(k_1,k_2,k_3)\rightarrow  V_a(k_1)~i {\bf W}_b^\a(k_2) D_\a
V_c(k_3)~\left[~{\rm Tr}(T^aT^bT^c)~ e^{-\frac{i}{2}(k_1
\times k_2+k_2\times k_3+k_1\times k_3)}\right.\nonumber\\
&&~~~~~~~~~~~~~~~~~~~~~~~~~~~~~~\left.-{\rm Tr}(T^cT^bT^a)~
e^{\frac{i}{2}(k_1 \times k_2+k_2\times k_3+k_1\times k_3)}\right]
\label{twistvert}
\eea
Now the quantum $V$ lines have  to be Wick
contracted in the consecutive order in which they appear.
As already mentioned, superspace $D$-algebra rules require two $D$ and two $\bar{D}$
spinor derivatives in the loop in order to obtain a nonzero
result and then the first
contribution to the effective action is from the four point function. From the Wick
expansion we look for
the forth order contribution
\beq
\frac{1}{4!(2g^2)^4}~({\cal U}+\bar{\cal U})^4
\label{four}
\eeq
and from (\ref{four}) we select the terms containing two ${\cal U}$'s
and two $\bar{\cal U}$'s.

There are two possible
arrangements of the vertices in the loop,
i.e.  ${\cal U}{\cal U}\bar{\cal U}
\bar{\cal U}$
and ${\cal U}\bar{\cal U}
{\cal U}\bar{\cal U}$.
For both cases the $D$-algebra is straightforward.
It gives $D_\a D_\b \Db_\ad \Db_\bd\rightarrow
C_{\b\a} C_{\bd\ad}$ and $D_\a \Db_\ad D_\b  \Db_\bd\rightarrow
-C_{\b\a} C_{\bd\ad}$ respectively for the two arrangements.
After $D$-algebra we can write the vertices symbolically as
(~see (\ref{twistvert})~)
\bea
&&{\cal U}^\a(k_1,k_2,k_3)
\equiv {\cal U}_P^\a(k_1,k_2,k_3)+{\cal U}_T^\a(k_1,k_2,k_3)
\equiv
V_a(k_1)~i {\bf W}_b^\a(k_2)
V_c(k_3)~\nonumber\\
&&~~~~~\nonumber\\
&&~~~~~~~~\left[~{\rm Tr}(T^aT^bT^c)~ e^{-\frac{i}{2}(k_1
\times k_2+k_2\times k_3+k_1\times k_3)}-{\rm Tr}(T^cT^bT^a)~
e^{\frac{i}{2}(k_1
\times k_2+k_2\times k_3+k_1\times k_3)}\right]\nonumber\\
&&~~~~~
\label{newvertex}
\eea
again with the $V$
quantum lines to  be contracted in the  order in which they appear in
the loop. We have found it convenient to write the external
background fields in the most symmetric way, that is
\bea
&&{\cal T}(1a,2b,3c,4d)={\bf W}^{\a a}(p_1){\bf W}_\a^b(p_2)
\bar{{\bf W}}^{\ad c}(p_3)
\bar{{\bf W}}_\ad^d(p_4)~~~~~~~~~~~~~~~~~~~~~~~~\nonumber\\
&&~~~~~~~~~~~~~~~~~~~~~~~+{\bf W}^{\a a}(p_1)\bar{{\bf W}}^{\ad b}(p_2)
\bar{{\bf W}}_\ad^c(p_3)
{\bf W}_\a^d(p_4)\nonumber\\
&&~~~~~~~~~~~~~~~~~~~~~~~-{\bf W}^{\a a}(p_1)\bar{{\bf W}}^{\ad b}(p_2)
{\bf W}_\a^c(p_3)
\bar{{\bf W}}_\ad^d(p_4)~+~{\rm h.c.}
\label{symmetric}
\eea
where the minus sign in the third term takes into account the result
from the $D$-algebra.
The above expression is completely symmetric in the exchanges of any
couple $(1a)\leftrightarrow (2b)\leftrightarrow (3c) \leftrightarrow
(4d)$, a property that we will use extensively in order to write the final
result in a simple form. Moreover from (\ref{symmetric}) one
obtains the corresponding bosonic expression in a form which is
directly comparable with the result from string amplitude
calculations
\bea
&&\int d^2\theta ~d^2\bar{\theta}~\left[{\bf W}^{\a a}(p_1){\bf W}_\a^b(p_2)
\bar{{\bf W}}^{\ad c}(p_3)
\bar{{\bf W}}_\ad^d(p_4)\right.~~~~~~~~~~~~~~~~~~~~~~~~\nonumber\\
&&~~~~~~~~~~~~~~~~~~~~~~~+{\bf W}^{\a a}(p_1)\bar{{\bf W}}^{\ad b}(p_2)
\bar{{\bf W}}_\ad^c(p_3)
{\bf W}_\a^d(p_4)\nonumber\\
&&~~~~~~~~~~~~~~~~~~~~~~~\left.-{\bf W}^{\a a}(p_1)\bar{{\bf W}}^{\ad b}(p_2)
{\bf W}_\a^c(p_3)
\bar{{\bf W}}_\ad^d(p_4)~+~{\rm h.c.}\right]\nonumber\\
&&\rightarrow t^{\a\b\g\d\m\n\rho\s} F_{\a\b}^a(p_1)F_{\g\d}^b(p_2)
F_{\m\n}^c(p_3)F_{\rho\s}^d(p_4)
\label{bosonic}
\eea
where $t^{\a\b\g\d\m\n\rho\s}$ is the symmetric tensor given
e.g. in formula (9.A.18) of \cite{GSW}.

\vskip 34pt
\noindent
\begin{minipage}{\textwidth}
\begin{center}
\includegraphics[width=0.40\textwidth]{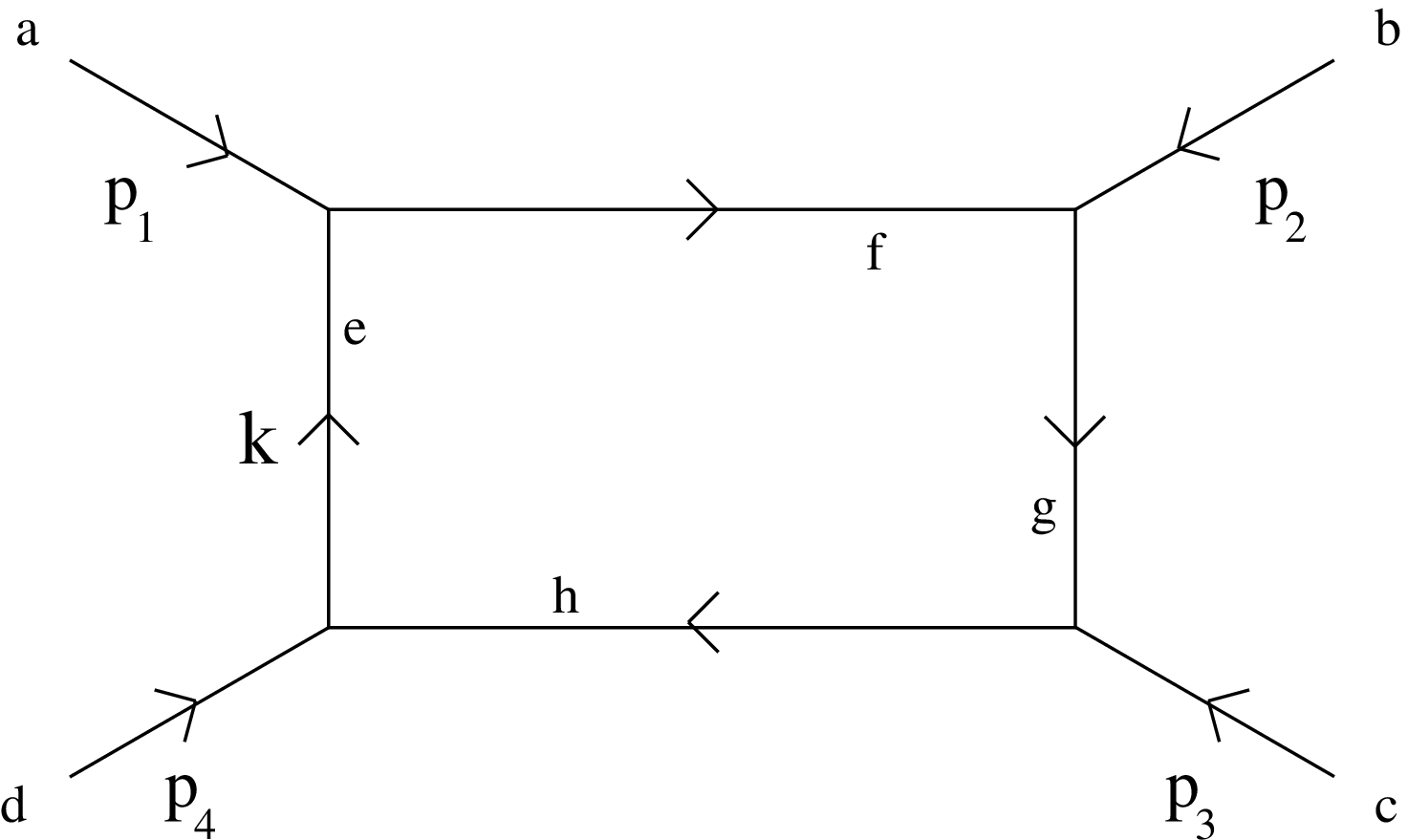}
\end{center}
\begin{center}
{\small{Figure 1:
box diagram}}
\end{center}
\end{minipage}

\vskip 36pt
\noindent
\begin{minipage}{\textwidth}
\begin{center}
\includegraphics[width=0.40\textwidth]{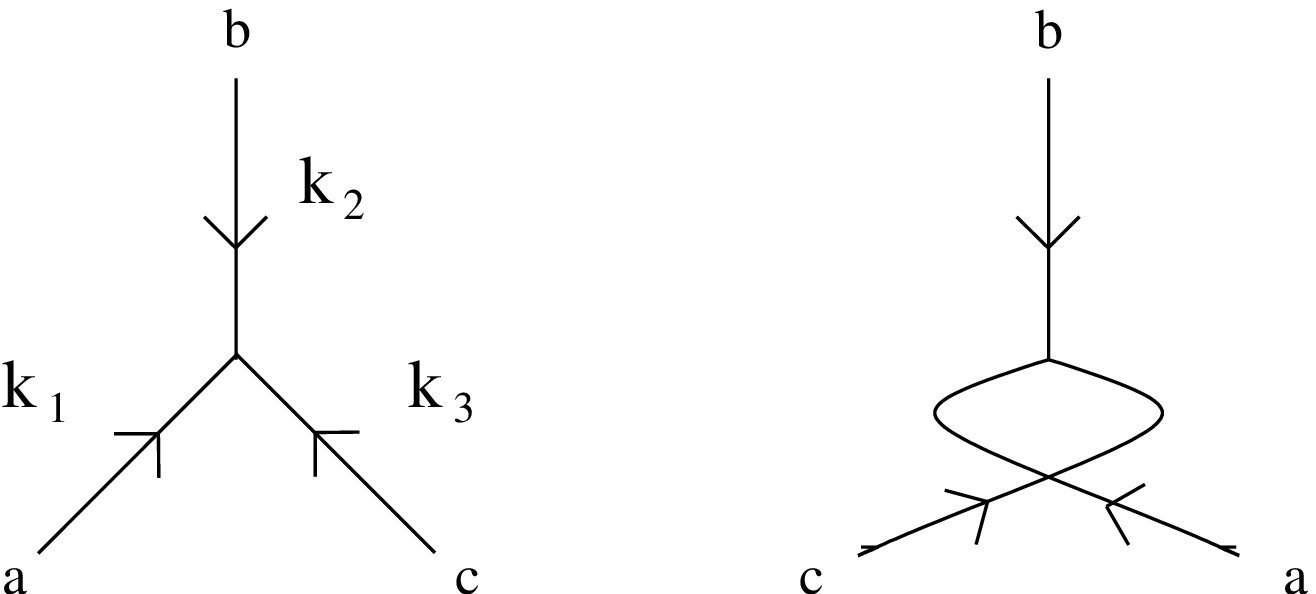}
\end{center}
\begin{center}
{\small{Figure 2:
untwisted and twisted vertices}}
\end{center}
\end{minipage}

\vskip 34pt

With all of this in mind the box diagram
can be written as
\beq
\left[~{\cal U}^\a~{\cal U}_\a~ \bar{\cal U}^\ad~ \bar{\cal
U}_\ad~+~{\cal U}^\a~ \bar{\cal U}^\ad~ \bar{\cal
U}_\ad~{\cal U}_\a~ -~{\cal U}^\a~ \bar{\cal U}^\ad~ {\cal U}_\a~
\bar{\cal U}_\ad+~{\rm h.c.}~\right]
\label{box}
\eeq
with ${\cal U}^\a$ given in (\ref{newvertex}). Each term
in (\ref{box}) produces
sixteen contributions: two of them, i.e. the ones which contain all
untwisted $P $ and all twisted $T $ vertices correspond to planar diagrams.
All the others, i.e. the ones with two $P$ and two $T$ vertices (a
total of six), the ones with one $P$ and three $T$'s (a total of four)
and the ones with one $T$ and three $P$'s (a total of four), correspond
to nonplanar graphs. In any case contraction of the quantum $V$'s
produces four scalar propagators as in (\ref{prop}). If we call $k$
the loop momentum and $p_1$, $p_2$, $p_3$, $p_4$ the external
momenta as shown in Fig. 1, we have
\bea
I_0(k;p_1,\dots,p_4)&=&
\frac{1}{(k+p_1)^2 k^2
(k-p_4)^2 (k+p_1+p_2)^2} \nonumber\\
&=&\int_0^\infty~ \prod_{i=1}^4 ~d\a_i~
e^{-\a\{k+\frac{1}{\a}[\a_1 p_1-\a_3 p_4
+\a_4(p_1+p_2)]\}^2}\nonumber\\
&&~~~~~~~~~e^{-[\a_1 p_1^2 +\a_3 p^2_4+ \a_4(p_1+p_2)^2]}
~e^{\frac{1}{\a}[\a_1 p_1-\a_3p_4+\a_4(p_1+p_2)]^2}
\label{4prop}
\eea
where $\a=\a_1+\a_2+\a_3+\a_4$.
The vertices can be either of the {\em untwisted} $P$ or of the {\em
twisted} $T$ type. With a labeling as in Fig. 2, we have
\bea
&&P\rightarrow {\rm Tr}(T^aT^bT^c)~ e^{-\frac{i}{2}(k_1
\times k_2+k_2\times k_3+k_1\times k_3)}={\rm Tr}(T^aT^bT^c)~
 e^{-\frac{i}{2}k_2\times k_3}\nonumber\\
&&T\rightarrow -{\rm Tr}(T^cT^bT^a)~
e^{\frac{i}{2}(k_1
\times k_2+k_2\times k_3+k_1\times k_3)}=-{\rm Tr}(T^cT^bT^a)~
e^{\frac{i}{2}k_2\times k_3}
\label{PandT}
\eea
where we have used $k_1+k_2+k_3=0$. For the diagram in Fig. 1 we have
two of the following choices at each vertex
\bea
&&P_1={\rm Tr}(T_eT_aT_f)~ e^{\frac{i}{2}p_1\times k }
 \qquad\quad ~~~~~{\rm or}\quad\qquad
  T_1=-{\rm Tr}(T_f T_aT_e)~ e^{-\frac{i}{2}p_1\times k }\nonumber\\
&&P_2={\rm Tr}(T_fT_bT_g)~ e^{\frac{i}{2}p_2\times(k+p_1)  }
\qquad\quad {\rm or} \quad\qquad T_2=-{\rm Tr}(T_gT_bT_f)~
 e^{-\frac{i}{2}p_2\times(k+p_1)  }
\nonumber\\
&&P_3={\rm Tr}(T_gT_cT_h)~ e^{\frac{i}{2}p_3\times(k-p_4)  }
\qquad\quad {\rm or}\quad\qquad
T_3=-{\rm Tr}(T_hT_cT_g)~ e^{-\frac{i}{2}p_3 \times (k-p_4)  }\nonumber\\
&&P_4={\rm Tr}(T_hT_dT_e)~ e^{\frac{i}{2}p_4\times k }
\qquad\quad~~~~~ {\rm or}\quad\qquad
T_4=-{\rm Tr}(T_eT_dT_h)~ e^{-\frac{i}{2}p_4\times k }
\label{allPT}
\eea

In the following two sections we will compute the full
contribution to the effective
action arising from all the possible
combinations of untwisted and twisted vertices.

\section{Planar contributions}

The planar diagrams correspond to terms  which
contain either four untwisted $P$ vertices or four twisted
$T$  vertices. In these cases the internal momentum $k$ cancels from
the exponential factors in (\ref{allPT}), and the remaining
dependence from the external momenta
is exactly such to reconstruct the $*$-product
among the ${\bf W}$'s \cite{DZ}.
By using the relation
$T^a_{ij}T^a_{kl}=\d_{il}\d_{jk}$ for the $U(N)$ gauge matrices, we
obtain the following contribution to the effective action \cite{DZ}
\bea
&&\G_{{\rm planar}}=\frac{N}{64} {\rm Tr}~ \int d^2\theta~d^2\bar{\theta}~
\frac{d^4p_1~d^4p_2~d^4p_3~d^4p_4}{(2\p)^{16}}~\d(\sum
p_i)~\int d^4k~I_0(k;p_1,\dots, p_4)~~~\nonumber\\
&&~~~~\nonumber\\
&&~~
\left( {\bf W}^\a(p_1)*{\bf W}_{\a }(p_2)*
 \bar{{\bf W}}^\ad(p_3)* \bar{{\bf W}}_{\ad }(p_4)+
{\bf W}^\a(p_1)* \bar{{\bf W}}^\ad(p_2)* \bar{{\bf W}}_{\ad }(p_3)*
{\bf W}_{\a }(p_4)
\right.\nonumber\\
&&~~~~~\nonumber\\
&&~~~~~~~~~~~~\left.-
{\bf W}^\a(p_1)*\bar{{\bf W}}^\ad(p_2)*{\bf W}_{\a }(p_3)*
\bar{{\bf W}}_{\ad }(p_4)~+~{\rm h.c.}
\right)
\label{boxplanar}
\eea
with $I_0$ defined in (\ref{4prop}).
Making use of the relation in (\ref{bosonic}) one obtains
the corresponding bosonic expression.
The result in (\ref{boxplanar}) is no surprise: it corresponds to the
result in the commutative theory with ordinary products replaced by
$*$-products.

\section{Non planar contributions}

Now we focus on the analysis of nonplanar contributions which represent
the real novelty as compared to standard perturbative
calculations. Indeed the phases from the $*$-products at the
vertices end up containing a dependence on the loop momentum; this
fact drastically changes the behaviour of the loop integral
\cite{perturb}.

The various nonplanar
diagrams group themselves into  two  classes \cite{DZ},  graphs
with two twisted vertices and graphs with one (or equivalently three)
twisted vertex.
For the first type of diagrams the trace on the
$U(N)$ matrices gives a factor like ${\rm Tr}(T^p T^q){\rm
Tr}(T^rT^s)$, while for the second type it gives ${\rm Tr}( T^p) {\rm Tr}(T^q
T^rT^s)$.
The total result can be written in the following form
\bea
&&\G_{{\rm nonplanar}}= \frac{1}{64}
\int d^2\theta~d^2\bar{\theta}~
\frac{d^4p_1~d^4p_2~d^4p_3~d^4p_4}{(2\p)^{16}}~\d(\sum
p_i)~{\cal T}(1a,2b,3c,4d)\nonumber\\
&&~~\int d^4k~\int_0^\infty~ \prod_{i=1}^4 ~d\a_i~
e^{-\a k^2}~
e^{-\frac{1}{\a}f(p_1,p_2,p_3,p_4;\a_i)}
~\left[ ~{\cal A}(k,p_1,\dots,p_4)~ +~{\cal B}(k,p_1,\dots,p_4)~\right]
\nonumber\\
&&~~~~~
\label{nonplanartotal}
\eea
where ${\cal T}(1a,2b,3c,4d)$ is defined in (\ref{symmetric}) and we
have performed the shift
\beq
k\rightarrow k-\frac{1}{\a}[\a_1 p_1-\a_3 p_4
+\a_4(p_1+p_2)]
\label{kshift}
\eeq
in the loop-momentum integral. Moreover we have defined the symmetric
function
\bea
f(p_1,p_2,p_3,p_4;\a_i) &=&
p_1^2 \left[ \a_1\a_2 + \frac14 (\a_2\a_4 + \a_1\a_3) \right]
+ p_2^2 \left[ \a_1\a_4 + \frac14 (\a_2\a_4 + \a_1\a_3) \right]
\nonumber\\
&+&  p_3^2 \left[ \a_3\a_4 + \frac14 (\a_2\a_4 + \a_1\a_3) \right]
+ p_4^2 \left[ \a_2\a_3 + \frac14 (\a_2\a_4 + \a_1\a_3) \right]
\nonumber\\
&+& \frac12~(p_1\cdot p_2 + p_3\cdot p_4 - p_1\cdot p_4 - p_2\cdot p_3)
~(\a_2\a_4 - \a_1\a_3)
\nonumber\\
&-& \frac12~(p_1\cdot p_3 + p_2\cdot p_4)
~(\a_2\a_4 + \a_1\a_3)
\label{fsimm}
\eea
With ${\cal A}(k,p_1,\dots,p_4)$ and ${\cal
B}(k,p_1,\dots,p_4)$ we have denoted the sum of all the contributions
proportional to  ${\rm Tr}(T^p T^q){\rm
Tr}(T^rT^s)$ and to ${\rm Tr}( T^p) {\rm Tr}(T^q
T^rT^s)$ respectively. We list now separately the terms in ${\cal A}$
and  ${\cal B}$.\\
\vspace{0.3cm}

\noindent
We have
\bea
&&{\cal A}(k,p_1,\dots,p_4)=P_1P_2T_3T_4+P_1T_2T_3P_4
+P_1T_2P_3T_4~~~~~~~~~~\nonumber\\
&&~~~~~~~~~~~~~~~~~~~~~~~+T_1 T_2P_3P_4+T_1P_2P_3T_4+T_1P_2T_3P_4
\label{calA}
\eea
where
\bea
&&P_1P_2T_3T_4 \rightarrow {\rm Tr}(T^a T^b){\rm Tr}(T^cT^d)~
e^{-\frac{i}{2}(p_1\times p_2-p_3\times p_4)}\nonumber\\
&&~~~~~~~~~~~~~~~~~e^{-i k\times(p_1+p_2)}~e^{\frac{i}{\a}[\a_1
p_1\times p_2-\a_3 p_3\times p_4]}
\nonumber\\
&&~~~~~~\nonumber\\
&&P_1T_2T_3P_4\rightarrow {\rm Tr}(T^a T^d){\rm Tr}(T^bT^c)~
e^{-\frac{i}{2}(p_1\times p_4-p_2\times p_3)}\nonumber\\
&&~~~~~~~~~~~~~~~~~e^{-i k\times(p_1+p_4)}~e^{\frac{i}{\a}[(\a_1+\a_3+\a_4)
p_1\times p_4-\a_4 p_2\times p_3]}\nonumber\\
&&~~~~~~\nonumber\\
&&P_1T_2P_3T_4\rightarrow {\rm Tr}(T^a T^c){\rm Tr}(T^bT^d)~
e^{-\frac{i}{2}(p_1\times p_3-p_2\times p_4)}\nonumber\\
&&~~~~~~~~~~~~~~~~~e^{-i k\times(p_1+p_3)}~e^{\frac{i}{\a}[(\a_1+\a_4)
p_1\times p_3-(\a_3+\a_4) p_2\times p_4]}
\label{2P2T}
\eea
The additional three contributions $T_1 T_2P_3P_4$, $T_1P_2P_3T_4$
and $T_1P_2T_3P_4$ are obtained from the corresponding terms in
(\ref{2P2T}) by hermitian conjugation.\\
\vspace{0.3cm}

\noindent
In the same way we have
\bea
&&{\cal B}(k,p_1,\dots,p_4)=P_1P_2P_3T_4+P_1T_2P_3P_4+P_1P_2T_3P_4+
T_1P_2P_3P_4~~~~~~~~\nonumber\\
&&~~~~~~~~~~~~~~~~~~~~~~~~+T_1T_2T_3P_4+T_1P_2T_3T_4+T_1T_2P_3T_4
+P_1T_2T_3T_4
\label{calB}
\eea
where
\bea
&&P_1P_2P_3T_4\rightarrow ~-~{\rm Tr}(T^a T^bT^c){\rm Tr}(T^d)
e^{-\frac{i}{2}(p_1\times p_2+p_2\times p_3+p_1\times
p_3)}\nonumber\\
&&~~~~~~~~~~~~~~~~~~~~e^{-ik\times(p_1+p_2+p_3)}
 ~e^{\frac{i}{\a}[\a_1 p_1\times p_2+\a_4p_2\times
 p_3+(\a_1+\a_4)p_1\times p_3]}
\nonumber\\
&&~~~~\nonumber\\
&&P_1T_2P_3P_4\rightarrow ~-~{\rm Tr}(T^cT^dT^a){\rm Tr}(T^b)
e^{-\frac{i}{2}(p_1\times p_3+p_3\times p_4+p_1\times
p_4)}\nonumber\\
&&~~~~~~~~~~~~~~~~~~~~e^{-ik\times(p_1+p_3+p_4)}
 ~e^{\frac{i}{\a}[(\a_1+\a_4) p_1\times p_3+\a_3p_3\times
 p_4+(\a_1+\a_3+\a_4)p_1\times p_4]}
\nonumber\\
&&~~~~\nonumber\\
&&P_1P_2T_3P_4\rightarrow ~-~{\rm Tr}(T^dT^aT^b){\rm Tr}(T^c)
e^{-\frac{i}{2}(p_1\times p_2+p_2\times p_4+p_1\times
p_4)}\nonumber\\
&&~~~~~~~~~~~~~~~~~~~~e^{-ik\times(p_1+p_2+p_4)}
 ~e^{\frac{i}{\a}[\a_1 p_1\times p_2+(\a_3+\a_4)p_2\times
 p_4+(\a_1+\a_3+\a_4)p_1\times p_4]}
\nonumber\\
&&~~~~\nonumber\\
&&T_1P_2P_3P_4\rightarrow ~-~{\rm Tr}(T^bT^cT^d){\rm Tr}(T^a)
e^{-\frac{i}{2}(p_2\times p_3+p_3\times p_4+p_2\times
p_4)}\nonumber\\
&&~~~~~~~~~~~~~~~~~~~~e^{-ik\times(p_2+p_3+p_4)}
 ~e^{\frac{i}{\a}[\a_4 p_2\times p_3+\a_3p_3\times
 p_4+(\a_3+\a_4)p_2\times p_4]}
\label{3PT}
\eea
As in the previous case the additional terms $T_1T_2T_3P_4$,
$T_1P_2T_3T_4$, $T_1T_2P_3T_4$ and $P_1T_2T_3T_4$ are the hermitian
conjugates of the contributions in (\ref{3PT}).

Now in (\ref{nonplanartotal}) the presence of
the completely symmetric expression ${\cal T}(1a,2b,3c,4d)$ allows to
freely symmetrize the rest of the terms with respect to the exchanges
$(1a)\leftrightarrow (2b)\leftrightarrow (3c) \leftrightarrow (4d)$.
Operating in this way we rewrite
\bea
&&e^{-\frac{1}{\a}f(p_1,p_2,p_3,p_4;\a_i)}
{\cal A}(k,p_1,\dots,p_4)  \rightarrow {\rm Tr}(T^a T^b)~{\rm Tr}(T^cT^d)
~e^{-\frac{i}{2}(p_1\times p_2-p_3\times p_4)}
~e^{-i k\times(p_1+p_2)}\nonumber\\
&&~~\nonumber\\
&&~~~~\left[~e^{-\frac{1}{\a}f(p_1,p_2,p_3,p_4;\a_i)}
e^{\frac{i}{\a}[\a_1 p_1\times p_2-\a_3 p_3\times p_4]}
+ e^{-\frac{1}{\a}f(p_1,p_3,p_4,p_2;\a_i)} e^{\frac{i}{\a}[(\a_1+\a_3+\a_4)
p_1\times p_2-\a_4 p_3\times p_4]}\right.\nonumber\\
&&~~~~\nonumber\\
&&~~~~~~~~~~~~~~~~\left. + e^{-\frac{1}{\a}
f(p_1,p_3,p_2,p_4;\a_i)}e^{\frac{i}{\a}[(\a_1+\a_4)
p_1\times p_2-(\a_3+\a_4) p_3\times p_4]}~\right]~+~{\rm h.c.}
\label{calAsymm}
\eea
Similarly we obtain
\bea
&&e^{-\frac{1}{\a}f(p_1,p_2,p_3,p_4;\a_i)}
{\cal B}(k,p_1,\dots,p_4)\rightarrow ~-~{\rm Tr}(T^a T^bT^c)~{\rm Tr}(T^d)
e^{-\frac{i}{2}(p_1\times p_2+p_2\times p_3+p_1\times
p_3)}\nonumber\\
&&~~~~\nonumber\\
&&~~~~~~e^{-ik\times(p_1+p_2+p_3)}~
\left[~ e^{-\frac{1}{\a}f(p_1,p_2,p_3,p_4;\a_i)}
e^{\frac{i}{\a}[\a_1 p_1\times p_2+\a_4p_2\times
p_3+(\a_1+\a_4)p_1\times p_3]}\right.\nonumber\\
&&~~~~~\nonumber\\
&&~~~~~~~~~+ e^{-\frac{1}{\a}f(p_1,p_4,p_2,p_3;\a_i)}
e^{\frac{i}{\a}[(\a_1+\a_4) p_1\times p_2+\a_3p_2\times
p_3+(\a_1+\a_3+\a_4)p_1\times p_3]}\nonumber\\
&&~~~~~\nonumber\\
&&~~~~~~~~~+ e^{-\frac{1}{\a}f(p_1,p_2,p_4,p_3;\a_i)}
e^{\frac{i}{\a}[\a_1 p_1\times p_2+(\a_3+\a_4)p_2\times
p_3+(\a_1+\a_3+\a_4)p_1\times p_3]}\nonumber\\
&&~~~~~\nonumber\\
&&~~~~~~~~~\left. + e^{-\frac{1}{\a}f(p_4,p_1,p_2,p_3;\a_i)}
e^{\frac{i}{\a}[\a_4 p_1\times p_2+\a_3p_2\times
p_3+(\a_3+\a_4)p_1\times p_3]}~\right] ~~+~{\rm h.c.}
\label{calBsymm}
\eea
Substituting (\ref{calAsymm}) and (\ref{calBsymm}) in
(\ref{nonplanartotal}) we obtain the full result for the one-loop
contributions to the four-point function of the ${\cal N}=4$
noncommutative Yang-Mills theory
\bea
&&\G_{{\rm total}}= \frac{1}{64}
\int d^2\theta~d^2\bar{\theta}~
\frac{d^4p_1~d^4p_2~d^4p_3~d^4p_4}{(2\p)^{16}}~\d(\sum p_i)~
~~~~~~~~~~~~~~~~~~~~~~~~~~~~~~~~~~~~~~~\nonumber\\
&&~~~~~~\left[{\bf W}^{\a a}(p_1){\bf W}_\a^b(p_2)
\bar{{\bf W}}^{\ad c}(p_3)
\bar{{\bf W}}_\ad^d(p_4)
+{\bf W}^{\a a}(p_1)\bar{{\bf W}}^{\ad b}(p_2)
\bar{{\bf W}}_\ad^c(p_3)
{\bf W}_\a^d(p_4)\right.\nonumber\\
&&~~~~~~~~~~~~~~~~~~~~~~~~~~~~\left.-{\bf W}^{\a a}(p_1)
\bar{{\bf W}}^{\ad b}(p_2){\bf W}_\a^c(p_3)
\bar{{\bf W}}_\ad^d(p_4)~+~{\rm h.c.}\right]
\nonumber\\
&&\int d^4k~\int_0^\infty~ \prod_{i=1}^4 ~d\a_i~
e^{-\a k^2}~\Big\{~N~{\rm Tr}(T^a T^b T^cT^d)
e^{\frac{i}{2}(p_2\times p_1 + p_1\times p_4 + p_2\times p_4)}
e^{-\frac{1}{\a}f(p_1,p_2,p_3,p_4;\a_i)}
\nonumber\\
&&~~~~~\nonumber\\
&&~~~~~~~~~~~~~~~~~~~+ ~{\rm Tr}(T^a T^b)~{\rm Tr}(T^cT^d)
~e^{-\frac{i}{2}(p_1\times p_2-p_3\times p_4)}
~e^{-i k\times(p_1+p_2)}\nonumber\\
&&~~~~\nonumber\\
&&\left[~ e^{-\frac{1}{\a}f(p_1,p_2,p_3,p_4;\a_i)}
e^{\frac{i}{\a}[\a_1 p_1\times p_2-\a_3 p_3\times p_4]}
+~ e^{-\frac{1}{\a}f(p_1,p_3,p_4,p_2;\a_i)} e^{\frac{i}{\a}[(\a_1+\a_3+\a_4)
p_1\times p_2-\a_4 p_3\times p_4]}\right.\nonumber\\
&&~~~~\nonumber\\
&&~~~~~~~~~~~~~~~~~~~~~~~~~~~~~~~~~~~~~~
\left.+ e^{-\frac{1}{\a}f(p_1,p_3,p_2,p_4;\a_i)} e^{\frac{i}{\a}[(\a_1+\a_4)
p_1\times p_2-(\a_3+\a_4) p_3\times p_4]}~\right]
\nonumber\\
&&~~~~\nonumber\\
&&~~~~~~~~~~~~~~~-~{\rm Tr}(T^a T^bT^c)~{\rm Tr}(T^d)
e^{-\frac{i}{2}(p_1\times p_2+p_2\times p_3+p_1\times
p_3)}~e^{-ik\times(p_1+p_2+p_3)}\nonumber\\
&&~~~~\nonumber\\
&&~~~~~~~~~~
\left[~ e^{-\frac{1}{\a}f(p_1,p_2,p_3,p_4;\a_i)}
e^{\frac{i}{\a}[\a_1 p_1\times p_2+\a_4p_2\times
p_3+(\a_1+\a_4)p_1\times p_3]}\right.\nonumber\\
&&~~~~\nonumber\\
&&~~~~~~~~~~
+~e^{-\frac{1}{\a}f(p_1,p_4,p_2,p_3;\a_i)}
e^{\frac{i}{\a}[(\a_1+\a_4) p_1\times p_2+\a_3p_2\times
p_3+(\a_1+\a_3+\a_4)p_1\times p_3]}\nonumber\\
&&~~~~\nonumber\\
&&~~~~~~~~~~
+~e^{-\frac{1}{\a}f(p_1,p_2,p_4,p_3;\a_i)}
e^{\frac{i}{\a}[\a_1 p_1\times p_2+(\a_3+\a_4)p_2\times
p_3+(\a_1+\a_3+\a_4)p_1\times p_3]}\nonumber\\
&&~~~~\nonumber\\
&&~~~~~~~~~~
\left. +~e^{-\frac{1}{\a}f(p_4,p_1,p_2,p_3;\a_i)}
e^{\frac{i}{\a}[\a_4 p_1\times p_2+\a_3p_2\times
p_3+(\a_3+\a_4)p_1\times p_3]}~\right]
~+~{\rm h.c.}~\Big\}
\label{totalsymm}
\eea
Using (\ref{bosonic}) one can extract the purely bosonic terms
contained in (\ref{totalsymm}).
The above result can be easily compared with corresponding
expressions obtained from the field theory limit of open string amplitudes
in the presence of a constant $B$ field. To this end it is sufficient
to express the Schwinger parameters $\a_i$, $i=1,\dots,4$,
in terms of variables $\l$, $\xi_i$, $i=1,2,3$, that
appear naturally in string loop calculations, e.g. with the ordering
$\xi_1>\xi_2>\xi_3$
\bea
\label{newvariables}
&&~~~~~~~~~~~~~~~~~~~~~~~~\l\equiv\a+\a_1+\a_2+\a_3+\a_4 \\
&& \a_1=\l (\xi_1-\xi_2)\qquad\qquad \a_2=\l(1-\xi_1)
\qquad\qquad \a_3=\l (\xi_2-\xi_3) \qquad\qquad \a_4=\l\xi_3
\nonumber
\eea
In the next section we make use of such changes of the integration variables
in order to study the four-point function in the low-energy limit.
In this case
 the integrations on the $\l$ and
$\xi_i$ variables can be performed exactly. Now we turn to this
computation.

\section{The low-energy four-point amplitude}

We study the amplitude (\ref{totalsymm}) in
the low-energy approximation $p_i\cdot p_j$ small,
with $p_i\times p_j$ finite.
In order not to worry about
IR divergences we use a mass regulator.
We introduce the change of variables in
(\ref{newvariables}) and perform the explicit integration
on the loop momentum $k$. We obtain
\bea
&&\G_{{\rm l.e.}} = \frac{\pi^2}{64}
\int d^2\theta~d^2\bar{\theta}~
\frac{d^4p_1~d^4p_2~d^4p_3~d^4p_4}{(2\p)^{16}}~\d(\sum p_i)~
~~~~~~~~~~~~~~~~~~~~~~~~~~~~~~~~~~~~~~~\nonumber\\
&&~~~~~~~~~~\left[{\bf W}^{\a a}(p_1){\bf W}_\a^b(p_2)
\bar{{\bf W}}^{\ad c}(p_3)
\bar{{\bf W}}_\ad^d(p_4)
+{\bf W}^{\a a}(p_1)\bar{{\bf W}}^{\ad b}(p_2)
\bar{{\bf W}}_\ad^c(p_3)
{\bf W}_\a^d(p_4)\right.\nonumber\\
&&~~~~~~~~~~~~~~~~~~~~~~~~~~~~~~~~~~\left.-{\bf W}^{\a a}(p_1)
\bar{{\bf W}}^{\ad b}(p_2){\bf W}_\a^c(p_3)
\bar{{\bf W}}_\ad^d(p_4)~+~{\rm h.c.}\right]
\nonumber\\
&&\int_0^1
d\xi_1\int_0^{\xi_1}d\xi_2\int_0^{\xi_2}d\xi_3
\int_0^\infty~d\l~\l
~\Big\{~e^{-\l m^2}~N~{\rm Tr}(T^a T^b T^cT^d)
~e^{\frac{i}{2}(p_2\times p_1 + p_1\times p_4 + p_2\times p_4)}
\nonumber\\
&&~~~~~\nonumber\\
&&~~~~~~~~+~e^{-\frac{1}{4\l}(p_1+p_2)\circ(p_1+p_2)-\l m^2}
~e^{-\frac{i}{2}(p_1\times p_2-p_3\times p_4)}
~{\rm Tr}(T^a T^b)~{\rm Tr}(T^cT^d)\nonumber\\
&&~~~~~\nonumber\\
&&~~~~~~~\left[~ e^{i[(\xi_1-\xi_2)
p_1\times p_2-(\xi_2-\xi_3) p_3\times p_4]} +~e^{i[\xi_1
p_1\times p_2-\xi_3 p_3\times p_4]}\right.\nonumber\\
&&~~~~\nonumber\\
&&~~~~~~~~~~~~~~~~~~~~~~~~~~~~~~~~~~~~~~~~~~~~
\left.+~e^{i[(\xi_1-\xi_2+\xi_3)
p_1\times p_2-\xi_2 p_3\times p_4]}~\right]
\nonumber\\
&&~~~~\nonumber\\
&&~~~-~e^{-\frac{1}{4\l}(p_1+p_2+p_3)\circ(p_1+p_2+p_3)-\l m^2}
~e^{-\frac{i}{2}(p_1\times p_2+p_2\times p_3+p_1\times p_3)}~
{\rm Tr}(T^a T^bT^c)~{\rm Tr}(T^d)
\nonumber\\
&&~~~~\nonumber\\
&&~~~~~~~~~~
\left[~ e^{i[(\xi_1-\xi_2) p_1\times p_2+\xi_3p_2\times
p_3+(\xi_1-\xi_2+\xi_3)p_1\times p_3]}+
e^{i[(\xi_1-\xi_2+\xi_3) p_1\times p_2+(\xi_2-\xi_3)p_2\times
p_3+\xi_1p_1\times p_3]}\right.\nonumber\\
&&~~~~\nonumber\\
&&~~~~~~~~~~
\left.+~e^{i[ (\xi_1-\xi_2)p_1\times p_2+\xi_2p_2\times
p_3+\xi_1p_1\times p_3]}
+e^{i[\xi_3 p_1\times p_2+(\xi_2-\xi_3)p_2\times
p_3+\xi_2p_1\times p_3]}~\right]\nonumber\\
&&~~~~~~~\nonumber\\
&&~~~~~~~~~~~~~~~~~~~~~~~~~~~~~~~~~~~~~~~~~~~~~~~~~~~~~~~~~~~~~
+~{\rm h.c.}~\Big\}
\label{lowenergytotal}
\eea
where we have defined $p\circ p=p_\m\Theta^{\m\n}\Theta_{\rho\n}p^{\rho}$.
Finally we perform the integration on the variables $\l$ and $\xi_i$.
Exploiting all the symmetries of the integrand
the complete result for the on-shell planar and nonplanar contributions
to the four-point function can be written as
\bea
&&\G_{{\rm l.e.}}=\frac{\pi^2}{64}
\int d^2\theta~d^2\bar{\theta}~
\frac{d^4p_1~d^4p_2~d^4p_3~d^4p_4}{(2\p)^{16}}~\d(\sum p_i)~
~~~~~~~~~~~~~~~~~~~~~~~~~~~~~~~~~~~~~~~\nonumber\\
&&~~~~~~~~~~\left[{\bf W}^{\a a}(p_1){\bf W}_\a^b(p_2)
\bar{{\bf W}}^{\ad c}(p_3)
\bar{{\bf W}}_\ad^d(p_4)
+{\bf W}^{\a a}(p_1)\bar{{\bf W}}^{\ad b}(p_2)
\bar{{\bf W}}_\ad^c(p_3)
{\bf W}_\a^d(p_4)\right.\nonumber\\
&&~~~~~~~~~~~~~~~~~~~~~~~~~~~~~~~~~~\left.-{\bf W}^{\a a}(p_1)
\bar{{\bf W}}^{\ad b}(p_2){\bf W}_\a^c(p_3)
\bar{{\bf W}}_\ad^d(p_4)~+~{\rm h.c.}\right]
\nonumber\\
&&~~~~\nonumber\\
&&~~~~~~\Big\{ ~\frac{1}{6m^4}~N~{\rm Tr}(T^a T^b T^c
T^d)~e^{\frac{i}{2}(p_2\times p_1+p_1\times p_4+p_2\times p_4)}
\nonumber\\
&&~~~~~~~~+~\frac{1}{m^2}~\frac{(p_1+p_2)\circ(p_1+p_2)}{4}
~\frac{\sin\left(\frac{p_1\times p_2}{2}\right)}
{\frac{p_1\times
p_2}{2}}~
\frac{\sin\left(\frac{p_3\times p_4}{2}\right)}{\frac{p_3\times
p_4}{2}}~\nonumber\\
&&~~~~\nonumber\\
&&~~~~~~~~~~~~~~~~~~~~~~~~~~~~
K_2(m\sqrt{(p_1+p_2)\circ(p_1+p_2)})
~{\rm Tr}(T^a T^b)~{\rm Tr}(T^cT^d)\nonumber\\
&&~~~~~\nonumber\\
&&~~~~~~~~-~\frac{1}{m^2}~\frac{(p_1+p_2+p_3)\circ(p_1+p_2+p_3)}{2}~
\frac{e^{-\frac{i}{2}(p_1\times p_2+p_2\times p_3+p_1\times p_3)}}
{(p_1\times p_4)(p_3\times p_4)}\nonumber\\
&&~~~~\nonumber\\
&&~~~~~~~~~~~~~~~~~~~~~~~~~~~~
K_2(m\sqrt{(p_1+p_2+p_3)\circ(p_1+p_2+p_3)})
~{\rm Tr}(T^a T^bT^c)~{\rm Tr}(T^d)
\nonumber\\
&&~~~~\nonumber\\
&&~~~~~~~~~~~~~~~~~~~~~~~~~~~~~~~~~~~~~~~~~~~~~~~~~~~~~~~~~~~
+~{\rm h.c.}~\Big\}
\label{lowenergyfinal}
\eea
Again using (\ref{bosonic}) we can extract the purely bosonic
contributions in terms of the $F_{\m\n}$ electromagnetic field
strengths. The resulting expression coincides with the one-loop field
theory limit obtained in \cite{*trek} from open string amplitudes.
We note that, as in \cite{*trek} only the terms from planar diagrams contain
phase factors depending on the momenta in such a way to reconstruct
the $*$-product between the superfield strengths.
The remaining two structures which arise from nonplanar contributions
have somewhat modified multiplication rules, which might endanger the
gauge invariance of the four-point amplitude.

\section{Conclusions}

The classical action in (\ref{SYMaction})
is invariant under nonlinear gauge transformations,
which are just the generalization to the noncommutative case
of the standard ones \cite{FL,superspace}
\bea
e^V&\rightarrow& e_*^{i\bar{\L}}*e_*^V*e_*^{-i\L}\nonumber\\
&&~~~~~~\nonumber\\
\Phi~\rightarrow~
e_*^{i\L}*\Phi* e_*^{-i\L}\qquad&&\qquad\bar{\Phi}~\rightarrow~
e_*^{i\bar{\L}}*\bar{\Phi}*e_*^{-i\bar{\L}}
\label{gaugetransf}
\eea
with a gauge parameter $\L$ which is a chiral superfield.

In the commutative case
the background covariant quantization is extremely advantageous for
gauge theories: the quantum-background splitting introduces separate
gauge transformations for the quantum and the background fields. The
gauge-fixing procedure breaks the quantum gauge invariance, while
maintaining  explicit the background gauge invariance. Thus  the
perturbative effective action is expressed in terms of covariant
derivatives and field strengths, with no need of implementing gauge
invariance through Ward identities.

When the noncommutative theories are constructed via the introduction
of the $*$-multiplication and the gauge transformations are modified
according to (\ref{gaugetransf}) the classical invariance is
guaranteed. Use of the background field method allows to obtain
perturbative results in terms again of covariant objects, but as we
have found in (\ref{lowenergyfinal}) the $*$-product is not
reproduced and new multiplication rules appear, i.e. $*'$- and
$*_3$-products \cite{*trek,gar,wise}.
Gauge invariance is not anymore manifest and
seems to be lost \cite{*trek,DZ}. One might worry that something went wrong
in applying standard perturbation theory techniques to a
theory which is noncommutative and non local. Or else should one
expect to recover gauge invariance only at the level of the
$S$-matrix?

It is worthwhile to investigate further.
The generalized $*$-products also appear in the solution of the
Seiberg-Witten equation \cite{SW,gar,wise} and in the expansion of gauge
invariant operators introduced in \cite{GHI,IKK}.
This seems to be an indication that
one is on the right track, even if not quite.
It has been shown that open Wilson lines play an important role
in the construction of gauge
invariant operators \cite{GHI,IKK} and it is conceivable that they might
be equally important at the level of the effective action. Following
this line of thoughts it
could be interesting to look for modifications of the Wick
 expansion prescription
or for a new mechanism that would allow to implement the open Wilson lines
directly in the action.
\vspace{1.5cm}

\noindent
{\bf Acknowledgements}

\noindent
This work has been partially supported by INFN, MURST and the
European Commission RTN program HPRN-CT-2000-00113 in which A.S. and D.Z.
are associated to the University of Torino.

\end{document}